\documentclass{emulateapj}
\usepackage{amssymb,amsmath,epsfig,graphicx, natbib}
\def\Hb{H$\beta$}
\def\Ha{H$\alpha$}
\def\Hd{H$\delta$}

\def\EWHB{W$_{H\beta}$}
\def\EWHD{W$_{H\delta}$}
\def\SFR{$\rm M_\odot ~yr^{-1}~kpc^{-2}$}
\def\A{\AA{}}

\begin{document}
\title{Stellar Population Gradients in ULIRGs: Implications for Gas Inflow Timescales}
\author{Kurt T. Soto, Crystal L. Martin}
\affil{Physics Department, University of California, Santa Barbara, CA 93106-9530}

\begin{abstract}
\label{sect:abst}
Using longslit, optical spectra of Ultraluminous Infrared Galaxies (ULIRGs), we measure the
evolution in the star-formation intensity during galactic mergers. In individual galaxies,
we resolve kpc scales allowing comparison of the nucleus, inner disk, and outer disk.
We find that the strength of the \Hb\ absorption line increases with the 
projected distance from the center of the merger, typically reaching about 9 \A around 10~kpc. At these radii, the
star formation intensity must have rapidly decreased about 300-400~Myr ago; only stellar populations deficient in 
stars more massive than Type~A produce such strong Balmer absorption. In contrast, we find
the star formation history in the central kpc consistent with continuous star formation. 
Our measurements indicate that gas depletion occurs from the outer disk inwards during major mergers.
This result is consistent with merger-induced gas inflow and empirically constrains the 
gas inflow timescale. Numerical simulations accurately calculate the total
amount of infalling gas but often assume the timescale for infall. 
These new measurements are therefore central to modeling merger-induced star formation and 
AGN activity. 

\end{abstract}

\subjectheadings{galaxies: starburst --- galaxies: evolution ---  galaxies: active --- galaxies: formation}

\section{Introduction}
\label{sect:intro}

Ultraluminous Infrared Galaxies (ULIRGs) are some of the most luminous objects in the local universe with $\log(L_{IR} / 
L_\odot) > 12$ -- a result of starburst and nuclear activity triggered by major mergers.  The number density of ULIRGs per 
unit comoving volume has evolved strongly causing ULIRGs to become a rare phenomenon locally, but infrared-luminous 
sources appear to dominate the star-forming activity beyond $z \sim 0.7$ \citep{lefloch05}.  By studying local 
analogues of these important objects we can better understand the physical processes regulating their evolution.

When stellar disks merge, phase-space density remains constant and therefore generally too low to match that of an elliptical galaxy
\citep{hernquist93}.  Merging gas-rich galaxies, however, allows dissipation; and the starburst increases the central 
phase-space density \citep{robertson06}. The extremely high star formation rates present in ULIRGs indicate high
central gas densities \citep{sanders86}, so ULIRGS meet the conditions for dissipative mergers. Further evidence that ULIRGs mark
this morphological transition comes from their stellar kinematics and surface brightness profiles, which resemble 
those of field elliptical galaxies \citep{genzel01}.  In the ULIRG phase, however, the gas mass is much higher
than that of an elliptical galaxy. The gas fraction blown away by stellar feedback is unclear, but half
or more is likely consumed by star formation. Where in the galaxy this star formation occurs is relevant to 
the steepening of the stellar age -- metallicity gradient and the fueling of the central supermassive black hole
in the merger remnant.

The high central concentration of molecular gas in ULIRGs indicates a strong burst of star formation in the central region 
of a kpc or less \citep{sanders86}. The rotation measured in some outflows, however, suggests a larger region
drives a galactic wind \citep{martin06}. In this paper, we use longslit, optical spectroscopy to examine radial gradients
in the stellar populations of ULIRGs. 
Integrated spectra, in contrast, are heavily weighted towards the recent 
star formation activity in the central kiloparsec. We present the sample in Section 2. In Section 3, we 
measure radial gradients in stellar spectral diagnostics.
We focus on the strength of Balmer absorption and emission. The hydrogen recombination rate directly measures the
ionization rate and is sensitive to the number of high-mass (and therefore young) stars. Balmer absorption is
strongest in Type~A stars, and high values indicate a paucity of OB stars and declining star formation rate.
In Section 4, we associate the cessation of star formation with depletion of the local gas density. We discuss
how our result constrains the gas inflow timescale and compare our measurement to the dynamical timescale.
Section 5 summarizes our conclusions.

\section{Data}

\subsection{Observations}

\label{sect:data}

The Keck~II echellete spectroscopy of 2~Jy~ULIRGs, described previously in \cite{martin05, martin06}, 
is well suited to our study of H Balmer emission/absorption. These data resolve scales
of 0.8\arcsec, about 1.5 kpc, spatially and $\sim70$~km~s$^{-1}$ spectrally. The 20\arcsec\ slit is long enough to sufficiently 
sample the sky at both ends of the slit, ensuring accurate sky subtraction and continuum intensity.  
The slit position angle includes a second nucleus when present (Table \ref{tab:class}). 
We inspected 43 ULIRG spectra and selected 25 for analysis of any stellar \Hb\ absorption.
This subsample presents \Hb\ emission such that the full width of the emission line in their nuclear spectrum, 
defined by the wavelengths where the trough changes from emission to absorption, is no more than  $1800\ km\ s^{-1}$.

The resulting subsample includes a range of stages in the merger progression.  
\cite{veilleux02} classified the merger stage for 11 of the 25 objects. Using K-band images from \cite{murphy96}, and
their identification as multiple nuclei systems, we apply the Veilleux scheme to the 
remainder finding 8 pre-mergers (i.e. two nuclei), a multiple merger  (IRAS 10565+2448), 12 mergers, and 4 old mergers.
The ULIRGS in the merger stage tend to present higher surface brightness, so our analysis of
spatial gradients in Balmer absorption is heavily weighted towards this merger phase.

\begin{deluxetable}{l c  c  l  c  c  c } 
\tabletypesize{\small}
\tablewidth{0pt}
\tablecaption{\label{tab:class}Object Morphology}
\tablehead{
\colhead{  IRAS name }	& \colhead{  IR  }& \colhead{  Opt }& \colhead{  Veil.  }& \colhead{  log($\frac{L_{IR}}{L_\odot}$)} & \colhead{  $z$} & \colhead{  PA}\\
 	(1)		& (2) & (3)  & (4)     & (5)                                               &  (6)  &  (7)\\
}
\startdata
17028+5817	&	D	&	D	&	 IIIa\tablenotemark{a}	&	12.38	&	0.1061	&	94.5		\\
23327+2913	&	D	&	D	&	 IIIa\tablenotemark{a}	&	12.03	&	0.1075	&	-4.3		\\
05246+0103	&	D	&	D	&	IIIb					&	12.05	&	0.0971	&	109.5	\\
00153+5454	&	D	&	D	&	IIIb					&	12.10	&	0.1116	&	-22.0		\\ 
16487+5447	&	D	&	D	&	 IIIb\tablenotemark{a}	&	12.12	&	0.1038	&	66.2		\\
16474+3430	&	D	&	D	&	 IIIb\tablenotemark{a}	&	12.12	&	0.1115	&	161.8	\\
15245+1019	&	D	&	St	&	 IIIb					&	11.96	&	0.0755	&	127.8	\\
20046-0623	&	S	&	St	&	 IIIb					&	12.02	&	0.0843	&	72.0		\\ 
20087-0308	&	S	&	St	&	 IVb					&	12.39	&	0.1057	&	85.5		\\ 
17574+0629	&	S	&	S	&	IVb					&	12.10	&	0.1096	&	51.2		\\ 
17208-0014	&	S	&	St	&	 IVb					&	12.38	&	0.0428	&	166.7	\\
18368+3549	&	S	&	S	&	IVb 					&	12.19	&	0.1162	&	-31.0		\\
23365+3604	&	S	&	St	&	 IVb					&	12.13	&	0.0645	&	-19.5		\\
09111-1007	&	S	&	W	&	 IVb					&	11.98	&	0.0542	&	73.5		\\ 
01298-0744	&	S	&	S	&	IVb\tablenotemark{a}	&	12.29	&	0.1362	&	-89.3		\\ 
11506+1331	&	S	&	S	&	IVb\tablenotemark{a}	&	12.27	&	0.1273	&	80.3		\\
10378+1109	&	S	&	S	&	IVb\tablenotemark{a}	&	12.23	&	0.1362	&	11.3		\\
11095-0238	&	S	&	St	&	IVb\tablenotemark{a}	&	12.20	&	0.1065	&	10.0		\\ 
18443+7433	&	S	&	St	&	 IVb					&	12.23	&	0.1347	&	29.6		\\
16090-0139	&	S	&	S	&	IVa\tablenotemark{a}	&	12.48	&	0.1336	&	107.6	\\ 
00262+4251	&	S	&	S	&	 V					&	12.08	&	0.0972	&	14.1		\\ 
08030+5243	&	S	&	S	&	 V					&	11.97	&	0.0835	&	0.0		\\ 
00188-0856	&	D	&	D	&	V\tablenotemark{a}		&	12.33	&	0.1285	&	-3.3		\\ 
10494+4424	&	S	&	S	&	 V					&	12.15	&	0.0923	&	25.0		\\
10565+2448	&	M	&	M	&	 M					&	11.98	&	0.0431	&	109.0	\\
\enddata

\tablenotetext{a}{Classifications directly from \cite{veilleux02}}
\tablecomments{ Col.(1): Name, Col.(2): Infrared K band morphology.  ``S'', ``D'', and ``M'' denote a single, double  and 
multiple nucleus morphology respectively \citep{murphy96}. Col.(3): Optical R band morphology.  Similar to R band 
morphology with an additional ``t'' to signify an extended tidal region, and ``W'' to denote a widely separated pair 
\citep{murphy96}. Col.(4): Merger Classification based on \cite{veilleux02}.  This classification separates wide binary pre-mergers 
(IIIa), close binary pre-mergers (IIIb), diffuse merger(IVa), compact merger (IVb), and old mergers (V). Col.(5): IR 
luminosity from \citep{murphy96}. Col.(6): redshift. Col.(7): Position angle of slit }

\end{deluxetable}

\subsection{Measurements}

\begin{figure}[h]
\centering
\epsfig{file=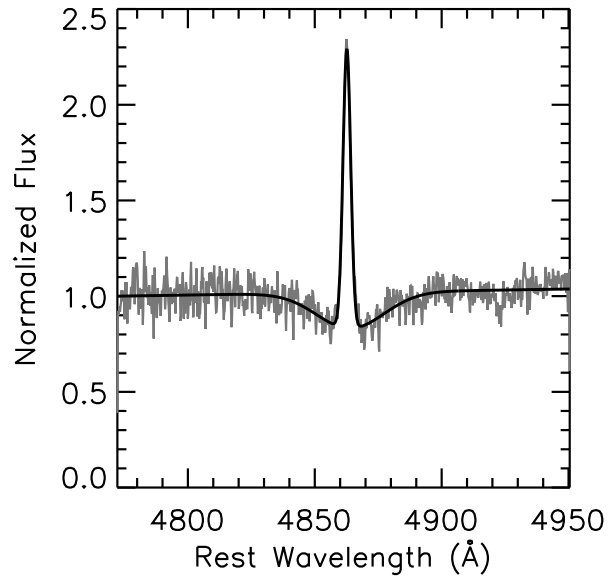, width=\linewidth}
\caption{\label{fig:hbprofile} The aperture selected at 2 kpc north west from the center in IRAS17208-0014 is shown above with a fit to both $H_\beta$ 
components simultaneously.  Emission component is a narrow line contributed by photoionization of the ISM.  The absorption component is provided 
by the presence of A stars in the stellar population.  The emission line is narrow enough that the absorption line is not affected in the fit.}
\vspace{0.5cm}
\end{figure}

\begin{figure}[h]
\centering
\epsfig{file=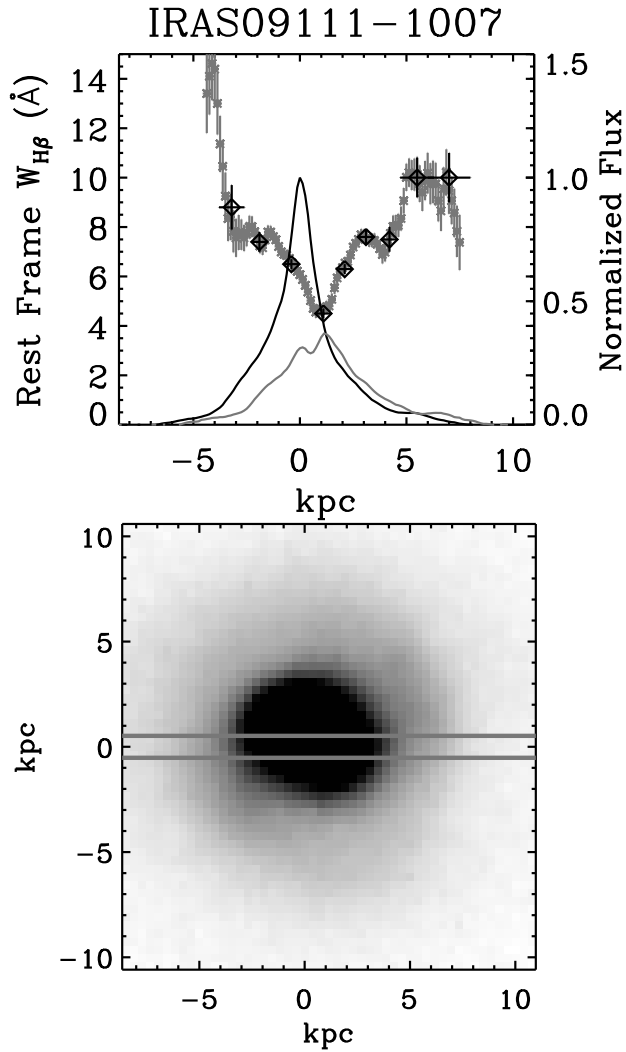, width=\linewidth}
\caption{\label{fig:ew_im}   
key : \textit{Upper:} The equivalent width of \Hb\ due to stellar absorption is plotted versus position.  The light grey points 
are the measurements from the preliminary sliding aperture analysis of the spectrum.  Black points are chosen
 apertures based on trends in the preliminary analysis.  The solid black line shows continuum intensity near \Ha\ as a
  function of position in units normalized to the peak of the continuum, illustrating the alignment with the images in the 
  lower panel and an indication of the relative signal to noise in the extended regions.  The solid grey line is the same for 
  the continuum near \Hb.  \textit{Lower:} An R band image (courtesy of T. Murphy) is shown with the position of the slit 
  plotted to show where the 1" $\times$ 20" slit was placed. We can see increasing \EWHB\ in the extended regions with a 
  sharp drop in the nucleus. Figures 2.a -- 2.j are available in the online version of the Journal. }
\vspace{0.5cm}
\end{figure}

We  examined the positional dependence of the \Hb\ line profile using a sliding aperture for spectral extraction.
We matched the aperture width to a seeing element ($\sim0''.8$) and extracted a spectrum at each line of the spectrogram.
At some spatial positions,  as illustrated in Fig. \ref{fig:hbprofile},  the spectra present  a two-component \Hb\  line 
profile. A broad absorption trough  underlies the narrow emission line.  We fit the absorption and emission components 
simultaneously using non-linear least-squares fitting with MPFIT in IDL.
These spectra exhibit continuous change from position to position due to atmospheric smearing, so parameters found from fitting one aperture position 
were supplied as the guess for the fit of the next position. The small, gray points in Figure \ref{fig:ew_im} show the result. Over scales of a few kpc,
this analysis often reveals an increase in \Hb\ absorption equivalent width, \EWHB, with distance from the peak continuum emission in the R band.

\begin{figure}[h]
\centering
\epsfig{file=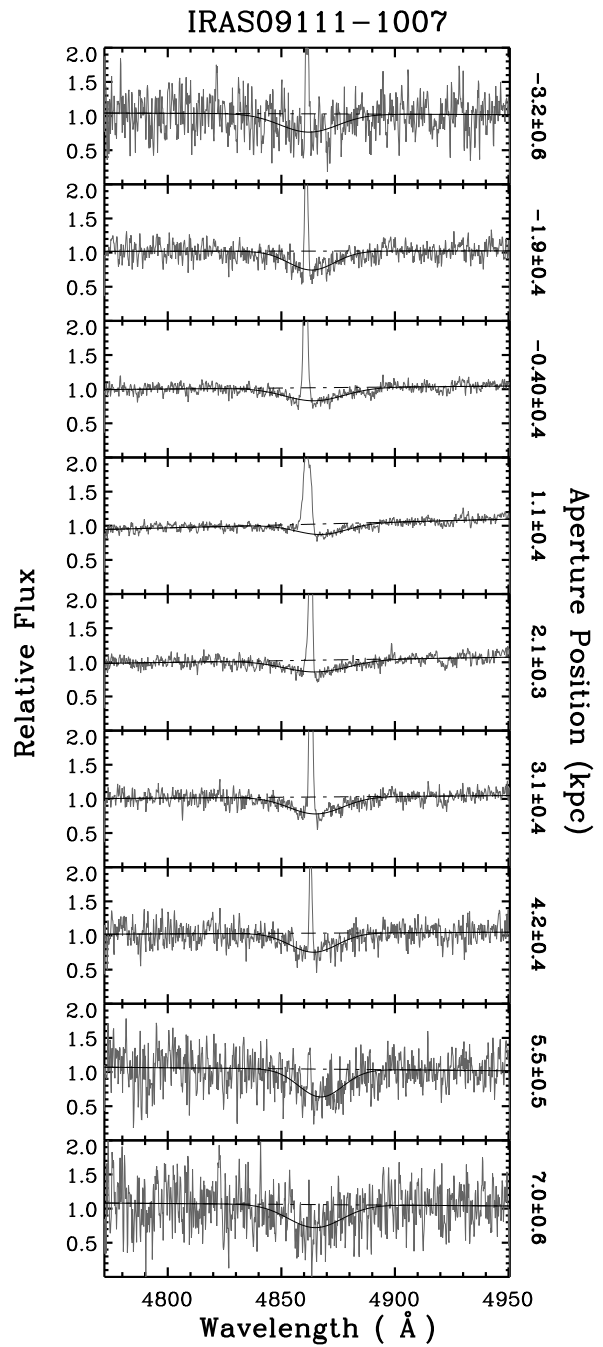,width=\linewidth}
\caption{\label{fig:stack_09111}  The extracted spectra used for measurement are plotted in grey with the fit to the
 absorption overplotted in black.  Each of these spectra correspond to one of the black points in Figure \ref{fig:ew_im}. Figures 3.a -- 3.j are included in the online version.}
\vspace{0.5cm}
\end{figure}

To eliminate effects due to correlated errors, we extracted spatially separated spectra based on the shape of these radial \Hb\ absorption  
and continuum surface brightness trends. These spectra are shown in Figure \ref{fig:stack_09111}, where the larger black 
points in Figure  \ref{fig:ew_im} show the \Hb\ absorption equivalent width obtained by fitting these spectra. 
We measured \Ha\ surface brightness in these same apertures. 
The \Hd\ lines could only be fitted in more than 3 apertures for one object.

To correct the \Ha\ emission for underlying stellar \Ha\ absorption, we measured  
the \Hb\ absorption equivalent width  and assumed the \Ha\ absorption equivalent width was two-thirds as large. 
Inspection
of synthetic spectra computed for stellar population models  \citep{bc03} verified that this scaling works over a broad
range of star formation models, including those with truncated star formation histories. The maximum value of the correction, 
3\A,  is small compared to the emission equivalent width.
The flux in \Ha\ emission is also corrected for extinction.
We determined the color excess, E(B -- V), per aperture by measuring the Balmer decrement and using a standard 
reddening law (R$_V$ = 4.05 = A$_V$/E(B -- V)) and the extinction law from \cite{calzetti00}. 
When significant, we discuss the correction for attenuation by Galactic dust on a galaxy-by-galaxy basis
in Section \ref{sect:spat_res}.  
We estimate areal star formation rate using the conversion $SFR({\rm M_\odot~yr^{-1}~kpc^{-2}}) =  [L(H\alpha)/A({\rm kpc^2})] / 1.26~\times~10^{41}~{\rm erg~s^{-1}}$ \citep{kennicutt98}.

\section{Results}

\begin{figure}[h]
\centering
\epsfig{file=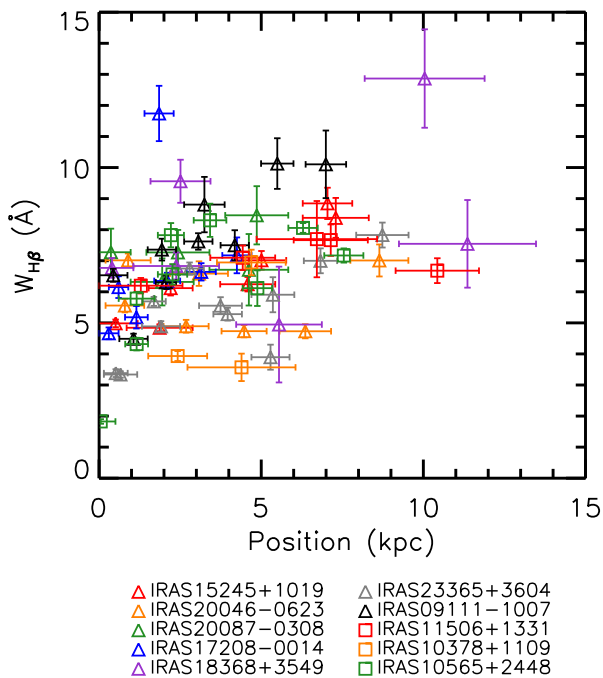,width=\linewidth}
\caption{\label{fig:all_ew} Equivalent width of \Hb\ absorption is plotted as a function of distance from the maximum
 continuum surface brightness.  The absorption strength generally increases away from the center.}
\end{figure}

The presence of \Hb\ emission and absorption reveal different epochs of star formation. 
Recombination emission traces ionization by very massive stars, and Balmer absorption is strongest in Type~A stars.
Strong Balmer absorption dominates the stellar spectrum when hotter stars are absent. 
Figure \ref{fig:all_ew} shows how the measured \Hb\ absorption equivalent width changes along
the slit. The projected distance is measured from the position
along the slit with the highest R band surface brightness.  This {\it center}
typically coincides with the position of the nucleus measured in the K band.
IRAS17208-0014 has two maxima in R band continuum, however only the brightest of these maxima coincides with a K-band nucleus, which we regard as the center of the object.
IRAS15245+1019 has two nuclei resolved in K-band images \citep{murphy96} separated by $\sim$ 4\arcsec, 
but we use the nucleus with the highest R-band surface brightness to define the center.  K-band imaging of IRAS20046-0623 
does not resolve two nuclei, but has elongated structure over 4\arcsec\ from east to west.  
The position of the maximum continuum surface brightness near \Ha\ differs from that of \Hb as well.  
For this object we chose to retain the definition of center based on the maximum R-band surface brightness since the K-band centroid is not
directly measured in these data.

The strength of the stellar Balmer absorption typically increases with distance 
from the center. For example, in Figure \ref{fig:all_ew} the \Hb\ absorption strength across IRAS 09111-1007
increases steadily out to 7 kpc. This behavior characterizes merger systems presenting
relatively symmetric structure on either side of the nucleus.
Objects such as IRAS20046-0623 observed at an earlier stage in the merger progression often
present more complicated morphology. The spatial gradient in \Hb\ equivalent width is not smooth like
that in IRAS09111-1007, but the absorption strength is still largest away from the center.
The surface brightness in the post mergers is generally too low to look for the effect at large radii.
The galaxies where such gradients can be measured fall in or between the close binary pre-merger stage (IIIa) to
compact merger stage (IVb). The 10 targets caught at  these stages form our spatially resolved subsample,
shown in Figure \ref{fig:all_ew} and discussed individually in Section~3.4.

\begin{figure}[h]
\centering
\epsfig{file=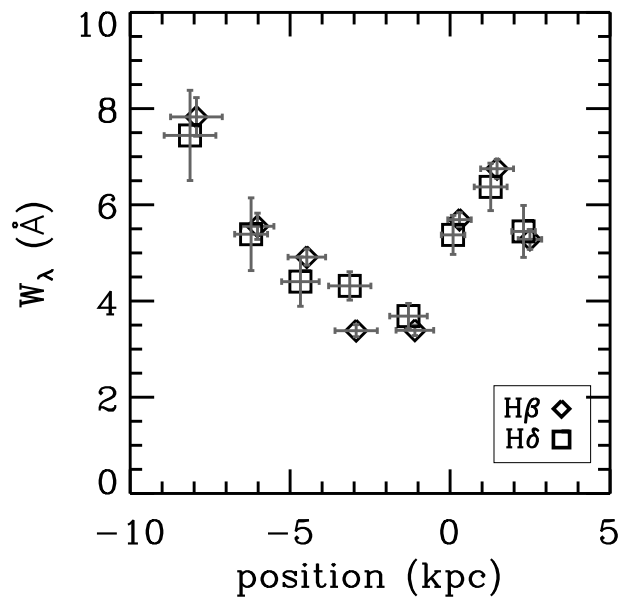,width=\linewidth}
\caption{\label{fig:multi_balmer} Equivalent widths of \Hb and \Hd\ absorption are plotted as a function of distance 
from the central continuum peak in the chosen apertures for IRAS23365+3604.  The Balmer equivalent widths 
appear to follow the same pattern.}
\end{figure}

Emission effects the line profile and equivalent width measurement less in \Hd.  However, IRAS 23365+3604 is the 
only galaxy where our data quality is sufficient to study their line profiles in more than 3 apertures. 
We fit the \Hb\ and \Hd\ emission/absorption profiles simultaneously, requiring the absorption features to 
have the same width and Doppler shift but allowing different normalizations.  
We apply the same requirement to the fit of the emission components.
Figure \ref{fig:multi_balmer} shows that the equivalent widths of these lines increases with 
distance from the center. Hence our results are not strongly influenced by our decomposition of the emission
and absorption components of the \Hb\ profile. We can use the measured strength
of Balmer absorption to constrain the recent star formation history at different radii.  The measurements in 
Figure \ref{fig:multi_balmer} also show that the equivalent width of \Hb\ and \Hd\ are similar in the apertures 
that could have both balmer lines measured.  The similarity is inconsistent with the population modeling which predicts 
slightly greater equivalent widths for \Hd.  Dominance of spectral types A0 through A5  could possibly explain this.  
A star formation history that allowed these stars to dominate might be required if measurements of more ULIRGs 
show this same inconsistency.

\subsection{Truncation Timescales from Stellar Population Synthesis}
\label{sect:sp_ages}

\begin{figure}[h]
\centering
	\epsfig{file=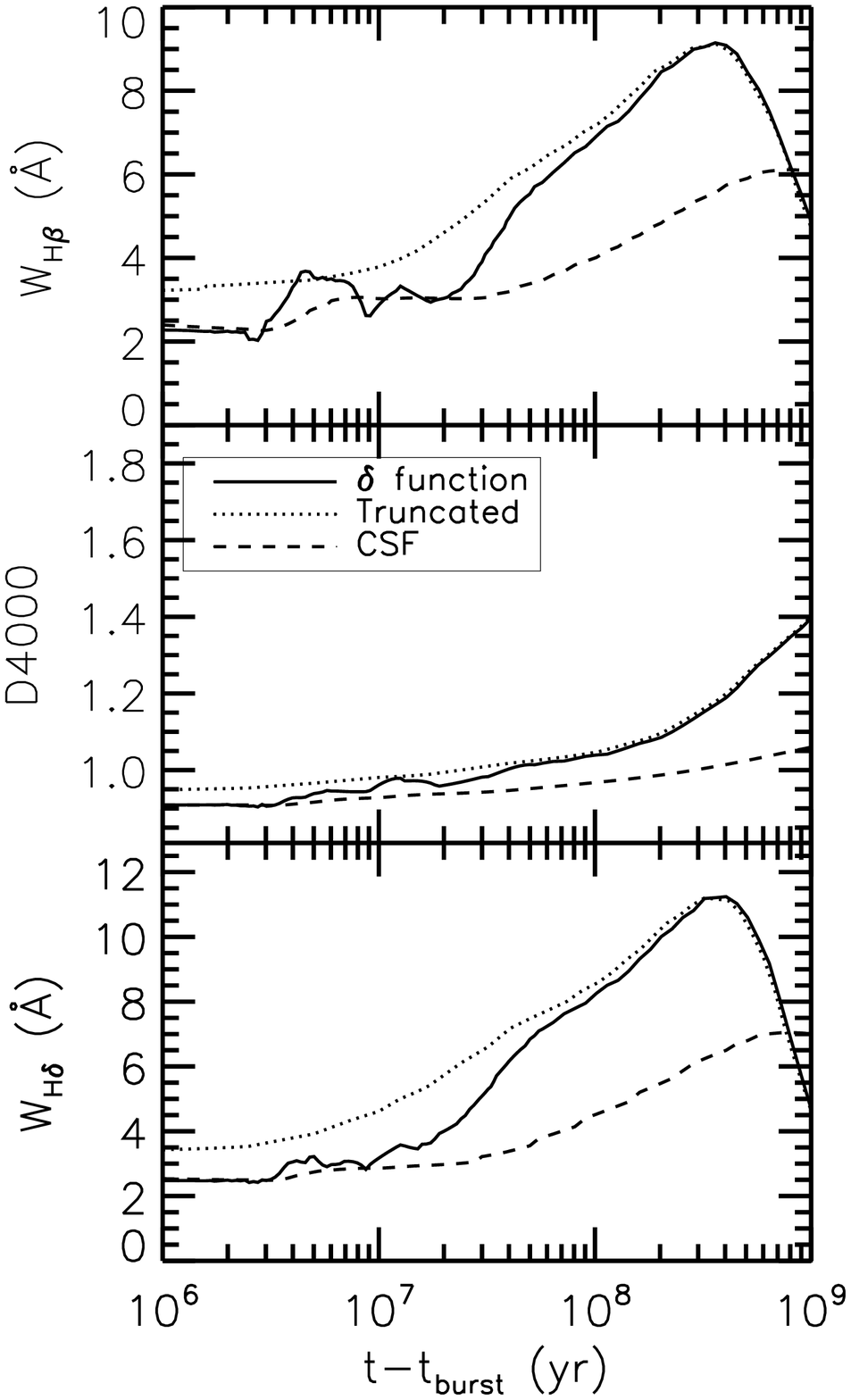,width=\linewidth, keepaspectratio=true}
\caption{\label{fig:agemeasures} Three stellar population age measures are compared using solar metallicity burst 
models from \cite{bc03}.  \EWHD shows slightly larger equivalent widths than \EWHB.  The 4000 Angstrom break (D4000) is defined by 
\cite{kauffmann03} as the ratio of fluxes in a bin redward of the break (4000 - 4100 \A) to a bin blueward of the break (3850 - 3950 \A).  
D4000 is monotonic and has an enhanced sensitivity over the range that the Balmer absorption function 
turns over creating an ambiguity in age.  Measurements of D4000 as a function of position for these objects will more 
clearly define population age.}
\end{figure}

Using GALAXEV \citep{bc03}, we synthesized model spectra of the stellar population
for three different star formation histories: (1) $\delta$ function, (2) continuous star formation history 
that is extinguished after 40 Myr, and (3) continuing star formation (CSF).   We assumed solar metallicities.
We measured \EWHB\ in these spectra using the same techniques applied to the observations. We illustrate
the evolution of the \EWHB\ as the stellar population ages in Figure \ref{fig:agemeasures}.
Around 850 Myr, continuous star formation reaches a maximum of 6 \A\  in equivalent width as the population reaches an 
equilibrium of births and deaths of A stars.  In the other models, as O and B type stars die off, the 
stellar population dominating the stellar continuum emission gradually shifts  to later stellar types.  
A single A5V star has \EWHB\ near 20 \A, but the spread in stellar types does not allow the composite models
to reach this limit. Once star formation is stopped, the \EWHB\ increases steadily towards a maximum of
 9\A\ at 400 Myr. After 400 Myr, the \EWHB\  falls as the number of A stars declines, reaching 6\A\ after
about 850 Myr. An equivalent width greater than 6 \A\ clearly requires a very rapid decrease in the 
star formation rate in the past Gyr. 

The large values of \EWHB\ found at radii $R > 5$~kpc in IRAS 23365+3604 are inconsistent 
with the continuous star formation rate model. Applying the relation between \EWHB\ and time since 
truncation, i.e. Figure \ref{fig:agemeasures} solid curve, we conclude that star formation 
ceased at least 100 Myr ago at $R > 5$~kpc.  The $H\delta$ equivalent width varies much like \Hb\ in 
Figure \ref{fig:agemeasures}  owing to its similar origin.
The \EWHB\ gradient continues towards the center of IRAS 23365+3604, and we interpret this as evidence that
the time since truncation decreases towards the center. In other words, star formation is apparently 
being shutdown from the outside inwards.

Another popular diagnostic of stellar age is the spectral break known as D4000. Our spectral bandpass 
does not completely cover this feature, but we can use the stellar population models along with our
observed trend in Balmer equivalent width to predict the radial variation in D4000. 
The D4000 index grows as the stellar population ages. Measurements of D4000 break the age degeneracy
to either side of the maximum in \EWHB\ at 400~Myr. Hence, we expect D4000 would show 
a steady increase with increasing radius in ULIRGs. 

Our simple stellar populations represent the most extreme limiting cases. A declining star formation
rate produces a rising \EWHB\ similar to our models with complete cessation. The stronger the rate of
change in the star formation rate, the closer the \EWHB\ comes to that of the truncated model. We 
can distinguish these two star formation histories by the absence/presence of Balmer emission in
our ULIRG spectra. The presence of nebular emission indicates the level of ongoing star formation.
Complete truncation of star formation activity reduces the ionization rate, and  Balmer emission
ceases. We examine the radial depedence of the implied star formation rate galaxy by galaxy.

\subsection{Spatially Resolved Star Formation Histories}
\label{sect:spat_res}

In Section 3, we identified 10 galaxies having extracted spectra in a least four apertures with SNR $\geq$ 
5.  Pre-merger and merger systems dominate this spatially resolved subsample.
  We discuss the recent star formation history in  each individually here.
For reference,  Table \ref{tab:posdep} lists the measured star formation rate and stellar population ages for different 
apertures in each object.  The star formation rates near the center should be treated cautiously due to two
important systematic errors.  First, an active galactic nucleus may contribute to the ionizing photon luminosity, thereby
lowering the inferred star formation rate. Second, the reddening measured by the Balmer decrement in the central
aperature reflects only the least obsured regions; the extinction may be too high in some regions for  Balmer
photons to escape causing us to underestimate the central star formation rate. These factors have little impact
on our analysis of spectra extracted outside of the central kiloparsec and our discussion of radial gradients.

 Our results suggest the following star formation histories.

\textit{IRAS15245+1019:}  This ULIRG has two K band nuclei separated by $\approx$ 4kpc.  These two nuclei host star 
formation of $4.83 \pm 0.05$ and $6.8 \pm 2.1$ \SFR\ for the brighter south-east nucleus and dimmer north west nucleus 
respectively.  The relative dimness in the visible band of the southeast nucleus is due to high levels of extinction by dust.  
The brighter nucleus in the interaction has \EWHB\ $5.0 \pm 0.1$ \A\ which indicates that star formation has been going 
on for $0.22 \pm 0.02$ Gyr. At 7~kpc, however, there is little to no star formation occurring currently. \EWHB\ indicates that 
$0.2 \pm 0.1$ Gyr has passed since truncation.  

\textit{IRAS20046-0623:}  From detailed analyses of galactic relative velocities and tidal tail characteristics, it is 
determined that IRAS20046-0623 has gone through pericenter in the last few times $10^{7}$ years \citep{murphy01}.  
The separated nuclei are masked within the extended central bulge that appears in both K band and R band imaging.  
The morphology of the extended tails are representative of two nearly orthogonal disks in the process of merging.  Peaks 
in star formation ($3.95 \pm 0.05$ and $8.3 \pm 0.7$ \SFR) occur in the nuclei indicated in \cite{murphy01}.  As shown in 
figure \ref{fig:ew_im}.c, these nuclear regions have \EWHB\ that indicate star formation has continued for 0.18 $\pm$ 0.03 
Gyr and 0.36 $\pm$ 0.07 Gyr.  In the outer regions $\sim$ 2 kpc beyond both nuclei, star formation is weaker, and their 
\EWHB\ is beyond that attainable by CSF.  The truncation time scale indicated by these measurements for these bins are 
from 80 to 100 Myr.  Galactic dust contributes $\sim$ 0.16 magnitudes of extinction at the redshifted wavelength of \Ha.  

\textit{IRAS20087-0308:}  The morphology of the K-band image shows a single nucleus, implying that the two 
galaxies have fully merged.  Similarly we see one centrally located position on the slit with continuing star 
formation.  The SNR for this object makes \EWHB\ difficult to measure, but the nuclear measurement in this object 
does appear slightly lower than the extended regions.  This difference is not significant enough to measure a 
gradient in the population age, so we estimate an overall age of 100 Myr since star formation stopped across the 
disk.  Galactic dust contributes $\sim$ 0.31 magnitudes of extinction at the redshifted wavelength of \Ha.

\textit{IRAS17208-0014:} This galaxy has a single K band nucleus with many diffuse disturbed regions around it in the R 
band.  The aperture with peak star formation ($0.64 \pm 0.03$ \SFR) is adjacent to the region with the smallest \EWHB\ 
($4.7 \pm 0.2$ \A ) which corresponds to $170 \pm 30$ Myr of continuous star formation (Figure \ref{fig:ew_im}.e).  Next 
to the K band centroid at the max of the R band continuum \EWHB\ = $6.6 \pm 0.3$ \A\ and the star formation is 
diminished to $0.016 \pm 0.003$ \SFR.  This aperture shows an underlying population with a truncation time scale of $ 
57 \pm 10 $ Myr.  The furthest aperture with a center 4.25 kpc from the nucleus has a slightly larger \EWHB\ and very little 
star formation, making the time since truncation of this population slightly larger at 100 $\pm$ 35 Myr.    Galactic dust 
contributes $\sim$ 0.67 magnitudes of extinction at the redshifted wavelength of \Ha. 

\textit{IRAS18368+3549:}  The K-band morphology is a single nucleus with the R band revealing a faint tidal region 
encircling the nucleus.  From this morphology we expect that the nuclei have completely merged and tidal distortions 
have had time to circularize.  The nucleus has an \EWHB\ of 6.8 $\pm$ 0.4 \A\ putting it beyond the maximum possible in 
CSF.  This measurement represents a time scale of $77 \pm 17$ Myr in the truncated constant star formation history, and 
$95 \pm 17$ Myr in the $\delta$-fn model.  \EWHB\ in the extended region at 10 kpc from the nucleus is very large putting 
it beyond the model describable with these star formation histories.

\textit{IRAS23365+3604:}  The morphology of this galaxy is similar to that of IRAS18368+3549, in that it has a bright 
nucleus with a tidal region that has had time to circularize.  The nuclear regions have a peak in star formation rate ($2.60 
\pm 0.02$ \SFR), and low \EWHB\ ($3.3 \pm 0.1$ \A).  We measure 45 $\pm$ 5 Myr since continuous star 
formation began.  Apertures exterior to this have similar times since a truncation of star formation ($37 \pm 3$ Myr in a 
truncated constant star formation history, $52 \pm 4$ Myr in a $\delta$-fn model).  At a distance of 5.3 kpc from the 
nucleus, the slit intersects an extended tidal region, the center of which shows a dip in equivalent width ($3.9 \pm 0.4$ \A) 
in an otherwise increasing \EWHB\ with radius.  In the furthest aperture no \Ha\ emission is measured, and the measured 
\EWHB\ of absorption indicates that truncation occurred $140 \pm 25$ Myr ago.
Galactic dust contributes $\sim$ 0.19 magnitudes of extinction at the redshifted wavelength of \Ha. 

\textit{IRAS09111-1007:} This galaxy has undisturbed tidal regions characteristic of post-merger morphology (Figure 
\ref{fig:ew_im}.a).  The central kpc has \EWHB\ = $ 4.5 \pm 0.2$ \A, which is indicative of younger stellar populations.  
From this point, \EWHB\ increases toward the edge of the object, up to 9 -10 \A\ at projected distances of 3 - 5 kpc.  These 
large equivalent widths imply that 300 Myr has passed since the truncation of star formation. 

\textit{IRAS10378+1109:} This galaxy has a compact nucleus, but it is measurable in several apertures.  The fit of the 
absorption trough in the central aperture yields an unphysically narrow fit due to contamination by strong emission and is rejected 
from the rest of the analysis.  This strong emission indicates strong star formation ($6.73 \pm 0.06$ \SFR).  The aperture 4.7 kpc 
has \EWHB\ = $ 6.9 \pm 0.4$ \A indicating a truncation time scale of $\sim 90$ Myr.

\textit{IRAS11506+1331:}  Star formation is maximum ($2.23 \pm 0.06$ \SFR) in the central most aperture (1.3 kpc).  If a continuing star 
formation history were considered, the absorption (\EWHB\ = $6.2 \pm 0.25$ \A) would indicate that constant star 
formation has continued for 1 Gyr, however the presence of a strong absorption line indicates that there was a previous 
burst of star formation that dominates the continuum over the stars being produced in the current epoch of star formation.  
In this case, the end of the previous burst of star formation would have ended $69 \pm 9$ Myr ago.  Exterior to this, 
regions have diminished star formation rates ($<$ 0.1 \SFR) and truncation time scales up to $150 \pm 35$ Myr at 7 kpc.  

\textit{IRAS10565+2448:}  This object is classified as a multiple merger by \citep{murphy96}, with a third component 
separated by 20 kpc from the brightest knot in our measurement.  These data sample two of the nuclei and measure an 
increase in \EWHB\ with radius.  In the stronger nucleus \EWHB\ drops to $1.8 \pm 0.1$ and has a large amount of star 
formation ($7.28 \pm 0.08$ \SFR ) indicating a very young population.  \EWHB\ increases to $8.3 \pm 0.5 $ \A, 
corresponding to a truncation time scale of $\sim 180$ Myr.  A dip in the trend occurs off center of the weaker nucleus.  
The truncation timescale in this region  is $\sim 50 Myr$.

From analysis of the spatially-resolved sub-sample, we draw the following conclusions.
Objects such as IRAS20087-0308, IRAS18368+3459, IRAS23365-3604 and IRAS10565+2448 have regions at large 
radii that show no \Ha, indicating that the star formation has been completely shut off. The others show examples of 
continuing, but decreased, star formation activity at larger radii.
Often the star formation rate per unit area drops by at least a factor of 10 at the largest radii measured.  Though the 
gradient varies among the galaxies, we see regularity in that the time since truncation increases outwards in radius.

\subsection{Other Sources of Spatial Gradients in \EWHB }

We considered other physical effects that could cause large \EWHB\ values.
Previous studies \citep{poggianti00} attributed large Balmer equivalent width to intermediate age stars having 
time to drift away from the dusty molecular clouds in which they were born.  This effect can have a major effect in 
integrated spectra of entire objects.  For dust to be the cause of larger equivalent width at larger radii, 
the population of hidden young stars must grow with distance from the nucleus. This conclusion contradicts
measurements of the molecular gas distribution in ULIRGs. An active nucleus may contribute to the continuum
emission in the central aperture. AGN typically contribute $15 -30 \%$ of the total bolometric luminosity from cool HII and LINER
ULIRGs \cite{veilleux09}. Boosting the continuum, however, only decreases the apparent stellar \EWHB, so an
AGN cannot explain large \EWHB\ values. Moreover, the gradients that we find on spatial scales larger than the seeing 
cannot be caused by an AGN.

\subsection{Comparison of \EWHB\ among Nuclei}
\label{sect:nuc}

\begin{figure}[h]
\centering
\epsfig{file=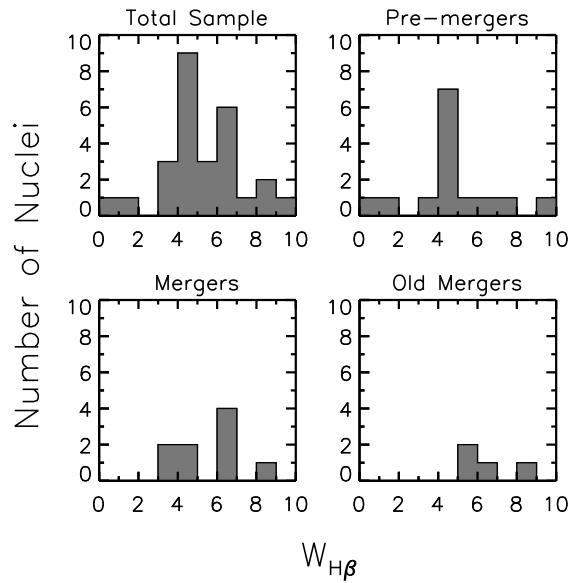,width=\linewidth}
\caption{\label{fig:hist} Histograms of the nuclear \EWHB\ measurements are presented for comparison between the 
different morphological classifications.  The pre merger histogram does not include 4 of the 18 nuclei that are present in 
the spectra due to the presence of emission broad enough to make an absorption component can not be discerned.  The 
merger stage objects are predominantly between 4 and 7 \A.  The sample of old mergers is small, but the measured 
nuclei are greater than 5 \A.}
\end{figure}

Among the 25 spectra examined, the surface brightness of the merger remnant was not always high enough to measure the spatial 
dependance of the line profile.  We can,
however, compare \EWHB\ among all the nuclei as illustrated in Figure \ref{fig:hist}.   In the pre-merger subsample we 
measured \EWHB\ of both nuclei seen.  Of these 18 nuclei, 4 of them showed broad \Hb\ emission that did not allow the 
measurement of an underlying absorption line.  A typical nucleus in the pre-merger sample presents \EWHB\ of 4~\A, 
although the full range in \EWHB\ is large.   In the merger sample, the median \EWHB\ is higher. Our sample is too small 
to be certain that this difference is significant; but, if verified, would suggest the central star formation declines once the 
nuclei merge. In the post merger phase,
\EWHB\ would then continue to increase up to roughly 8 or 9~\A\ before starting to decline.

\section{Discussion}
\label{sect:discussion}

By resolving the recent star formation activity in ULIRGs over scales of 1-10 kpc, we learned that star formation is 
being turned off in ULIRGs from the outside inwards. Specifically, we measure an increase in \EWHB\ with increasing radius and
argue that the only reasonable interpretation is an increasing fraction of Type~A stars (relative to earlier types).
We go one step further here and interpret the reduced star formation rate as direct evidence for gas depletion. Above
a threshold density of approximately $\sim10~{\rm M_{\odot}~pc^{-2}}$, the star formation intensity empirically scales non-linearly with the gas surface 
density,   $\Sigma_{SFR} \propto \mu^{1.4}$, where $\mu$ is the surface density of gas  \citep{martin01,kennicutt98}. 
If star formation were the primary mechanism removing the gas, then the gas would be consumed more quickly in the highest density
regions, $\tau \sim \mu / SFR \propto \mu^{-0.4}$, i.e. the nuclei.  Star formation alone produces a gradient oppositely sloped 
compared to the one we measured. Our result therefore provides direct evidence that another mechanism moves gas out of the outer disk.
It is natural to ask whether merger-induced inflows are that mechanism.

Simulations of gas-rich, galaxy -- galaxy mergers predict an increase in star formation during the first passage and a 
second, stronger starburst
at the time of actual merger \citep{mihos94,lotz08, hopkins09}.  
The torques imparted on the gas during the merger cause the gas to flow inward, robbing the outer regions of fuel to 
generate stars.  The infall is a well known result of the separation of stars from gas in the merger remnant; a 
consequence of the different collisional properties of stars and gas \citep{hopkins09}.  This separation allows the stars to 
impart a torque on the gas and cause infall. 
Models imply the correlation between the dynamical age of the merger and the age of the starburst population \citep{lotz08}
but have yet to predict how much of the star formation occurs in the central kpc. 
A key aspect of merger models should be to trace gas migration and the location of star formation activity throughout the merger. 

Observations in \cite{kewley06b} provide circumstantial evidence relating gas inflow to the starburst and therefore to the interaction.  
Galaxy mergers with greater starburst strength have lower nuclear metallicity. This effect is expected to be due to the infall of pristine 
gas from the outer disk which dilutes the concentration of metals in the nucleus.  
We have measured the time of gas depletion at radii from 1 to 10 kpc. The results constrain the net infall velocity.

\subsection{Implication of Radial Gradients in Star Formation History for Gas Inflow Timescales}

Table \ref{tab:posdep} lists the time elapsed since the star formation rate or, equivalently, the gas surface density dropped abruptly. 
The time elapsed since peak star formation activity is typically 100 to 300 Myr at 5 to 10 kpc falling to less than 50 Myr within
5 kpc. These timescales indicate the minimum time for gas inflow since the gas need only flow inward, not necessarily  all the way to the 
center of the merger. 

In observations by \cite{colina05}, the dynamical mass for objects included in our study (IRAS17208+0014, IRAS20087-0308, IRAS 23365+3604) 
are measured as similar to the Milky Way mass.  The representation that we give for the orbital velocities may be reparameterized by substituting 
in distances and orbital velocities for other galaxies, as well as changing the density values used in the freefall calculation.  Observations 
\citep{dasyra06} have indicated that ULIRGs are major mergers between galaxies with an average mass ratio of 1.5:1, indicating that the 
orbital velocities in each component are of the same order.  This information allows us to choose the Milky Way mass value as the fiducial 
mass scale; and we estimate maximum inflow speeds of order $v_{inflow} \sim 68~{\rm km\ s^{-1}} (R/ 7 ~{\rm kpc}) / (t_* / 100~{\rm  Myr})$.

\begin{figure}[h]
\centering
\epsfig{file=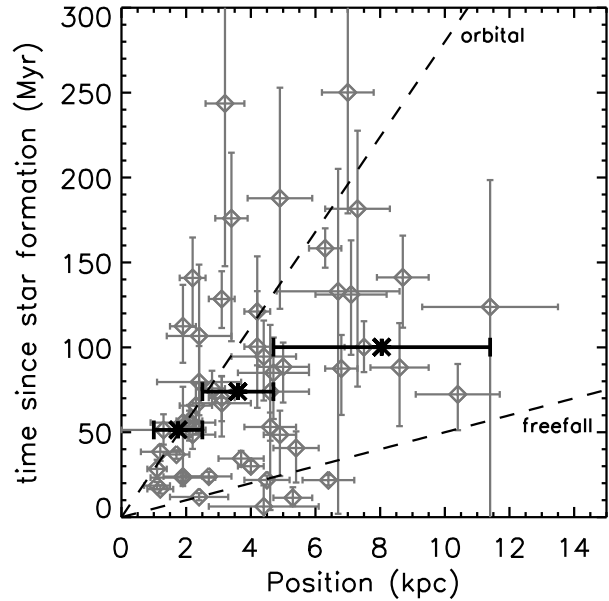}
\caption{\label{fig:pos_age} The grey points are the measured values for age and position in the truncated star formation 
history model.  The black points are the median values of age within bins that contain equal numbers of points, where the 
bin size is represented by the error bar.  The median points show that the infall timescale is on the same order as  the orbital 
period and the free fall time scale. The scatter at large radii may indicate differences in star formation history or 
inflow time scale.}
\end{figure}

 We compare this to two timescales in ULIRGs: (1) the free fall timescale at a given radius in an isothermal sphere 
 and  (2) the orbital period. Both of these timescales increase linearly with radius.  
To determine the freefall timescale we calculate the average density within particular radii using a density profile of the 
form $\rho_0\ (r/r_0)^{-2} $ with $\rho_0 = 1.3 \times 10^7~{\rm M_\odot~kpc^{-3}} $ the local halo mass density \citep{gates95}. 
The average density of the Milky Way interior to a radius of 8~{\rm kpc} is  $\rho_{MW} \approx 4 \times 10^7 ~{\rm M_\odot~kpc^{-3}}$ 
giving the free fall timescale $\tau_{ff}(r) \approx 40~{\rm Myr}~\sqrt{\rho_{MW}/\rho(r)}$.  The orbital period is  $\tau_{orb}(R) 
\approx 220 ~{\rm Myr}~ (R/8~{\rm kpc}) / (v / 220 ~{\rm km~ s^{-1}})$.  
In Figure  \ref{fig:pos_age}, we show the truncation timescale as a function of distance from the center of the
merger. The scatter in values does not distinguish between freefall and orbital time scales. They do rule out
the much longer time scales associated with diffusion and lend empirical support to the timescales assumed in
simulations \citep{hopkins08,hopkins09}.

\subsection{Picture Emerging from Comparison of Stellar and Dynamical Ages}

Gas migration toward the center slows down star formation in the outer radii. The paucity of massive stars allows \EWHB\ to grow.
Regions closer to the nucleus are fed infalling gas from exterior regions, prolonging star 
formation and allowing \EWHB\ to remain low longer.
As the merger advances, the cessation of star formation trails inward.
All 25 ULIRGs have star formation in their center, indicating the gas surface density there remains above the threshold density.
As discussed in Section \ref{sect:nuc}, the higher nuclear \EWHB\ in the merger and post-merger objects may indicate the central
star formation rate is declining by this stage. This result provides observational evidence that the central gas density 
starts declining once the nuclei have merged.

\section{Conclusion}
\label{sect:conclusion}

The physical processes regulating the rate of gas infall during mergers
is important because it influences the age and metallicity gradients in the merger remnant as well as the strength of supernova and AGN feedback.
By examining the positional dependence of stellar spectral indices in ESI long slit spectra , we determined the recent star formation history across
galaxy -- galaxy mergers. Strong \Hb\ absorption indicates a diminished star formation rate over the past few hundred Myr, which should be
accompanied by an increase in D4000, while hydrogen Balmer emission indicates the presence of massive stars from more recent star formation. 

We find the measured \EWHB\ increases 
with radius in the sample of spatially resolved objects. We attribute the large \EWHB\  to a rapid decrease in star formation activity
over 100~Myr ago at radii greater than 5~kpc. At radii of a few kpc, the activity appears to have decreased just 50 to 100 Myr ago based
on the slightly lower, but still prominent, \EWHB. Our measurements in the central kpc are consistent with star formation continuing unabated.
The nature of the gradient implies that gas was removed from larger radii first allowing the stellar populations to age.

We propose that these age gradients trace the inward flow of gas from large radii, adding to the central gas supply and prolonging the
central starburst. Our data primarily address the pre-merger and early merger phases. At later stages, the surface brightness of a ULIRG is
typically too low to make the equivalent measurement. We do, however, find preliminary evidence that these spectral indices in the nuclei of
late mergers indicate decreasing star formation. We interpret these results as direct evidence for strong gas inflow in the pre-merger
and early merger phases with a possible suppression by the late merger phase. 

Assuming the time since star formation diminished reflects the gas inflow timescale, our measurements indirectly 
constrain the gas infall rate.  The time scales for infall to occur are within the range of the orbital timescale and the freefall 
timescale of an isothermal sphere.  

\acknowledgments{
The authors thank Omer Blaes, Dawn Erb, Philip Hopkins, and Vivienne Wild for illuminating suggestions and comments.  
This work was supported by the National Science Foundation under contract 080816.  KS acknowledges additional support
from the Department of Education through the Graduate Assistance in Areas of National Need program.
The authors wish to recognize and acknowledge the very significant cultural role and reverence that the summit of Mauna Kea has always had 
within the indigenous Hawaiian community.  We are most fortunate to have the opportunity to conduct observations from this mountain.}

{\it Facilities:}  \facility{Keck}

\bibliographystyle{apj}
\bibliography{ulirg}

\begin{thebibliography}{25}
\expandafter\ifx\csname natexlab\endcsname\relax\def\natexlab#1{#1}\fi

\bibitem[{{Bruzual} \& {Charlot}(2003)}]{bc03}
{Bruzual}, G., \& {Charlot}, S. 2003, \mnras, 344, 1000

\bibitem[{{Calzetti} {et~al.}(2000){Calzetti}, {Armus}, {Bohlin}, {Kinney},
  {Koornneef}, \& {Storchi-Bergmann}}]{calzetti00}
{Calzetti}, D., {Armus}, L., {Bohlin}, R.~C., {Kinney}, A.~L., {Koornneef}, J.,
  \& {Storchi-Bergmann}, T. 2000, \apj, 533, 682

\bibitem[{{Colina} {et~al.}(2005){Colina}, {Arribas}, \&
  {Monreal-Ibero}}]{colina05}
{Colina}, L., {Arribas}, S., \& {Monreal-Ibero}, A. 2005, \apj, 621, 725

\bibitem[{{Dasyra} {et~al.}(2006){Dasyra}, {Tacconi}, {Davies}, {Genzel},
  {Lutz}, {Naab}, {Burkert}, {Veilleux}, \& {Sanders}}]{dasyra06}
{Dasyra}, K.~M., {et~al.} 2006, \apj, 638, 745

\bibitem[{{Gates} {et~al.}(1995){Gates}, {Gyuk}, \& {Turner}}]{gates95}
{Gates}, E.~I., {Gyuk}, G., \& {Turner}, M.~S. 1995, \apjl, 449, L123+

\bibitem[{{Genzel} {et~al.}(2001){Genzel}, {Tacconi}, {Rigopoulou}, {Lutz}, \&
  {Tecza}}]{genzel01}
{Genzel}, R., {Tacconi}, L.~J., {Rigopoulou}, D., {Lutz}, D., \& {Tecza}, M.
  2001, \apj, 563, 527

\bibitem[{{Hernquist} {et~al.}(1993){Hernquist}, {Spergel}, \&
  {Heyl}}]{hernquist93}
{Hernquist}, L., {Spergel}, D.~N., \& {Heyl}, J.~S. 1993, \apj, 416, 415

\bibitem[{{Hopkins} {et~al.}(2009){Hopkins}, {Cox}, {Younger}, \&
  {Hernquist}}]{hopkins09}
{Hopkins}, P.~F., {Cox}, T.~J., {Younger}, J.~D., \& {Hernquist}, L. 2009,
  \apj, 691, 1168

\bibitem[{{Hopkins} {et~al.}(2008){Hopkins}, {Hernquist}, {Cox}, {Dutta}, \&
  {Rothberg}}]{hopkins08}
{Hopkins}, P.~F., {Hernquist}, L., {Cox}, T.~J., {Dutta}, S.~N., \& {Rothberg},
  B. 2008, \apj, 679, 156

\bibitem[{{Kauffmann} {et~al.}(2003){Kauffmann}, {Heckman}, {Tremonti},
  {Brinchmann}, {Charlot}, {White}, {Ridgway}, {Brinkmann}, {Fukugita}, {Hall},
  {Ivezi{\'c}}, {Richards}, \& {Schneider}}]{kauffmann03}
{Kauffmann}, G., {et~al.} 2003, \mnras, 346, 1055

\bibitem[{{Kennicutt}(1998)}]{kennicutt98}
{Kennicutt}, Jr., R.~C. 1998, \apj, 498, 541

\bibitem[{{Kewley} {et~al.}(2006){Kewley}, {Geller}, \& {Barton}}]{kewley06b}
{Kewley}, L.~J., {Geller}, M.~J., \& {Barton}, E.~J. 2006, \aj, 131, 2004

\bibitem[{{Le Floc'h} {et~al.}(2005){Le Floc'h}, {Papovich}, {Dole}, {Bell},
  {Lagache}, {Rieke}, {Egami}, {P{\'e}rez-Gonz{\'a}lez}, {Alonso-Herrero},
  {Rieke}, {Blaylock}, {Engelbracht}, {Gordon}, {Hines}, {Misselt}, {Morrison},
  \& {Mould}}]{lefloch05}
{Le Floc'h}, E., {et~al.} 2005, \apj, 632, 169

\bibitem[{{Lotz} {et~al.}(2008){Lotz}, {Jonsson}, {Cox}, \& {Primack}}]{lotz08}
{Lotz}, J.~M., {Jonsson}, P., {Cox}, T.~J., \& {Primack}, J.~R. 2008, \mnras,
  391, 1137

\bibitem[{{Martin}(2005)}]{martin05}
{Martin}, C.~L. 2005, \apj, 621, 227

\bibitem[{{Martin}(2006)}]{martin06}
---. 2006, \apj, 647, 222

\bibitem[{{Martin} \& {Kennicutt}(2001)}]{martin01}
{Martin}, C.~L., \& {Kennicutt}, Jr., R.~C. 2001, \apj, 555, 301

\bibitem[{{Mihos} \& {Hernquist}(1994)}]{mihos94}
{Mihos}, J.~C., \& {Hernquist}, L. 1994, \apj, 427, 112

\bibitem[{{Murphy} {et~al.}(1996){Murphy}, {Armus}, {Matthews}, {Soifer},
  {Mazzarella}, {Shupe}, {Strauss}, \& {Neugebauer}}]{murphy96}
{Murphy}, Jr., T.~W., {Armus}, L., {Matthews}, K., {Soifer}, B.~T.,
  {Mazzarella}, J.~M., {Shupe}, D.~L., {Strauss}, M.~A., \& {Neugebauer}, G.
  1996, \aj, 111, 1025

\bibitem[{{Murphy} {et~al.}(2001){Murphy}, {Soifer}, {Matthews}, \&
  {Armus}}]{murphy01}
{Murphy}, Jr., T.~W., {Soifer}, B.~T., {Matthews}, K., \& {Armus}, L. 2001,
  \apj, 559, 201

\bibitem[{{Poggianti} \& {Wu}(2000)}]{poggianti00}
{Poggianti}, B.~M., \& {Wu}, H. 2000, \apj, 529, 157

\bibitem[{{Robertson} {et~al.}(2006){Robertson}, {Cox}, {Hernquist}, {Franx},
  {Hopkins}, {Martini}, \& {Springel}}]{robertson06}
{Robertson}, B., {Cox}, T.~J., {Hernquist}, L., {Franx}, M., {Hopkins}, P.~F.,
  {Martini}, P., \& {Springel}, V. 2006, \apj, 641, 21

\bibitem[{{Sanders} {et~al.}(1986){Sanders}, {Scoville}, {Young}, {Soifer},
  {Schloerb}, {Rice}, \& {Danielson}}]{sanders86}
{Sanders}, D.~B., {Scoville}, N.~Z., {Young}, J.~S., {Soifer}, B.~T.,
  {Schloerb}, F.~P., {Rice}, W.~L., \& {Danielson}, G.~E. 1986, \apjl, 305, L45

\bibitem[{{Veilleux} {et~al.}(2002){Veilleux}, {Kim}, \&
  {Sanders}}]{veilleux02}
{Veilleux}, S., {Kim}, D.-C., \& {Sanders}, D.~B. 2002, \apjs, 143, 315

\bibitem[{{Veilleux} {et~al.}(2009){Veilleux}, {Rupke}, {Kim}, {Genzel},
  {Sturm}, {Lutz}, {Contursi}, {Schweitzer}, {Tacconi}, {Netzer}, {Sternberg},
  {Mihos}, {Baker}, {Mazzarella}, {Lord}, {Sanders}, {Stockton}, {Joseph}, \&
  {Barnes}}]{veilleux09}
{Veilleux}, S., {et~al.} 2009, \apjs, 182, 628

\end{thebibliography}

\clearpage

\LongTables
\begin{deluxetable}{l l l l l l l}
\tablewidth{0pt}
\tablecaption{ \label{tab:posdep} Position Dependent Measurements}
\tabletypesize{\small}
\tablehead{
  \colhead{IRAS name} 	& \colhead{$r-r_{nuc}$}	&\colhead{ \EWHB}  & \colhead{CSF} & \colhead{TSFH} & \colhead{DSFH} & \colhead{SFR}    \\
&  (1) 	& (2) & (3)  & (4) & (5) & (6)     \\
}
\startdata
     15245+1019 &   -7.0 $\pm$  0.8 &    8.8  $\pm$  0.5 &                       N/A                    &  250.1  $^{+   71.2}_{-  71.2}$ &  263.6  $^{+   68.3}_{-  68.3}$ &   0.0014  $\pm$   0.0007\tablenotemark{a}   \\
                &   -5.0 $\pm$  0.8 &    7.0  $\pm$  0.3 &                       N/A                                	&   88.5  $^{+  21.0}_{-  14.3}$ &  106.8  $^{+  23.8}_{-  14.0}$ &   0.026  $\pm$   0.016 \tablenotemark{a}  \\
                &   -1.9 $\pm$  1.0 &    4.8  $\pm$  0.1 &  203.4  $^{+  13.8}_{-  20.7}$ 		&   23.3  $^{+   1.6}_{-   1.4}$ &   38.8  $^{+   1.1}_{-   0.7}$ &  1.52  $\pm$     0.02 \\
                &    0.5 $\pm$  0.7 &    5.0  $\pm$  0.1 &  221.3  $^{+  23.7}_{-  15.5}$ 		&   25.4  $^{+   1.9}_{-   1.8}$ &   40.3  $^{+   1.3}_{-   1.4}$ &  4.83  $\pm$     0.05 \\
                &    2.2 $\pm$  0.7 &    6.1  $\pm$  0.2 &  895.3  $^{+  404.4}_{- 404.4}$ 	&   48.6  $^{+   8.4}_{-   8.3}$ &   66.5  $^{+   8.3}_{-   7.2}$ &  1.01  $\pm$     0.05 \\
                &    4.6 $\pm$  0.8 &    6.2  $\pm$  0.2 &                       N/A                                	&   53.1  $^{+   8.1}_{-   8.5}$ &   70.9  $^{+   8.3}_{-   7.9}$ &  6.8  $\pm$     2.1 \\
                &    7.3 $\pm$  1.0 &    8.4  $\pm$  0.6 &                       N/A                                	&  181.6  $^{+ 104.7}_{-  46.0}$ &  197.8  $^{+ 105.6}_{-  41.6}$ &   0.008   $\pm$  0.006 \tablenotemark{a}  \\
                & & &  & & & \\ 

     20046-0623 &   -8.6 $\pm$  0.9 &    7.0  $\pm$  0.5 &                       N/A                    &   88.1  $^{+  34.5}_{-  26.2}$ &  106.5  $^{+  37.5}_{-  26.8}$ &  0.0053   $\pm$  0.0005 \\
                &   -6.4 $\pm$  0.8 &    4.7  $\pm$  0.2 &  182.0  $^{+  35.8}_{-  28.7}$ &   21.8  $^{+   3.1}_{-   3.0}$ &   38.0  $^{+   1.9}_{-   2.7}$ &  0.156  $\pm$     0.002 \\
                &   -4.5 $\pm$  0.7 &    4.7  $\pm$  0.2 &  182.7  $^{+  31.8}_{-  23.8}$ &   21.9  $^{+   2.7}_{-   2.6}$ &   38.1  $^{+   1.6}_{-   2.4}$ &   3.95  $\pm$     0.05 \\
                &   -2.7 $\pm$  0.7 &    4.9  $\pm$  0.2 &  209.6  $^{+  33.0}_{-  33.1}$ &   24.0  $^{+   3.2}_{-   2.7}$ &   39.2  $^{+   2.3}_{-   1.6}$ &  1.18  $\pm$     0.01 \\
                &   -0.8 $\pm$  0.6 &    5.6  $\pm$  0.2 &  360.0  $^{+  63.2}_{-  73.0}$ &   34.6  $^{+   3.8}_{-   3.7}$ &   50.4  $^{+   5.3}_{-   5.3}$ &  3.62  $\pm$     0.08 \\
                &    0.9 $\pm$  0.7 &    7.0  $\pm$  0.2 &                       N/A                                &   88.5  $^{+  16.1}_{-  10.5}$ &  106.8  $^{+  16.7}_{-  10.4}$ &  8.3  $\pm$     0.7 \\
                &    3.1 $\pm$  0.9 &    6.6  $\pm$  0.4 &                       N/A                                &   67.1  $^{+  19.6}_{-  15.8}$ &   85.3  $^{+  19.8}_{-  16.2}$ &   0.98  $\pm$     0.32 \\
                & & &  & & & \\ 

     20087-0308 &   -4.9 $\pm$  1.0 &    8.5  $\pm$  0.9 &                       N/A                                &  187.8  $^{+  65.1}_{-  65.1}$ &  204.1  $^{+  60.0}_{-  60.0}$ &   \tablenotemark{b}  \\
                &   -2.4 $\pm$  1.0 &    7.3  $\pm$  0.7 &                       N/A                                &  106.7  $^{+  46.7}_{-  42.0}$ &  127.7  $^{+  44.6}_{-  45.4}$               &   0.058  $\pm$   0.025\tablenotemark{a}  \\
                &   -0.4 $\pm$  0.6 &    7.3  $\pm$  0.8 &                       N/A                                &  107.1  $^{+  49.4}_{-  44.0}$ &  128.2  $^{+  46.9}_{-  47.6}$ &  \tablenotemark{d} \\
                &    1.9 $\pm$  1.0 &    6.3  $\pm$  0.8 &                       N/A                                &   55.7  $^{+  37.4}_{-  21.1}$ &   73.5  $^{+  36.5}_{-  23.1}$ &   0.8  $\pm$     0.2  \\
                &    4.6 $\pm$   1.2 &    6.7  $\pm$  1.2 &                       N/A                                &   73.9  $^{+  69.7}_{-  39.3}$ &   92.5  $^{+  70.9}_{-  42.1}$ &  0.04  $\pm$     0.05 \\
                & & &  & & & \\ 

     17208-0014 &   -4.2 $\pm$  0.4 &    7.2  $\pm$  0.6 &                       N/A                                &  100.4  $^{+  36.0}_{-  32.7}$ &  116.6  $^{+  40.2}_{-  30.6}$ &   0.004   $\pm$  0.003\tablenotemark{a}  \\
                &   -3.1 $\pm$  0.4 &    6.6  $\pm$  0.3 &                       N/A                                &   70.2  $^{+  13.7}_{-  12.9}$ &   89.0  $^{+  12.9}_{-  13.9}$ &  0.016  $\pm$     0.003 \\
                &   -2.1 $\pm$  0.4 &    6.4  $\pm$  0.3 &                       N/A                                &   56.8  $^{+  11.7}_{-   9.3}$ &   74.6  $^{+  12.4}_{-   9.2}$ &   0.099  $\pm$     0.004 \\
                &   -1.1 $\pm$  0.3 &    5.2  $\pm$  0.4 &  260.1  $^{+  94.3}_{-  60.8}$ &   28.4  $^{+   5.9}_{-   5.4}$ &   42.4  $^{+   7.8}_{-   3.7}$ &  0.52  $\pm$     0.02 \\
                &   -0.3 $\pm$  0.3 &    4.7  $\pm$  0.2 &  174.0  $^{+  30.5}_{-  25.1}$ &   20.9  $^{+   2.5}_{-   2.5}$ &   37.2  $^{+   1.6}_{-   2.2}$ &  0.51  $\pm$     0.01 \\
                &    0.6 $\pm$  0.3 &    6.2  $\pm$  0.4 &                       N/A                                &   49.8  $^{+  12.6}_{-  11.0}$ &   67.6  $^{+  12.4}_{-  10.8}$ &  0.64  $\pm$     0.03 \\
                &    1.8 $\pm$  0.4 &   12.0  $\pm$  0.9 &                       N/A                                &                       N/A                                &                       N/A                                &   0.11  $\pm$     0.02 \\
                & & &  & & & \\ 

     18368+3549 &  -10.0 $\pm$  1.8 &   12.9  $\pm$  1.6 &                       N/A                            &                       N/A                           &                       N/A                     &  \tablenotemark{b} \\
                &   -2.5 $\pm$  0.9 &    9.6  $\pm$  0.7 &                       N/A                              &                       N/A                              &                       N/A                            &   0.8   $\pm$  0.5 \tablenotemark{a}  \\
                &   -0.4 $\pm$  0.7 &    6.8  $\pm$  0.3 &                       N/A                                &   76.7  $^{+  17.1}_{-  16.3}$ &   95.2  $^{+  15.3}_{-  16.9}$ &  \tablenotemark{d} \\
                &    2.4 $\pm$  1.3 &    6.8  $\pm$  0.4 &                       N/A                                &   79.6  $^{+  26.9}_{-  21.2}$ &   97.9  $^{+  29.5}_{-  21.6}$ &  0.76  $\pm$     0.26 \\
                &   11.4 $\pm$  2.1 &    7.5  $\pm$  1.4 &                       N/A                                &  123.8  $^{+ 145.9}_{-  74.7}$ &  145.1  $^{+ 135.1}_{-  78.1}$ &   0.00008   $\pm$   0.00009 \tablenotemark{a}  \\
                & & &  & & & \\ 

     23365+3604 &   -8.7 $\pm$  0.8 &    7.8  $\pm$  0.4 &                      N/A                   	&  141.2  $^{+  29.6}_{-  24.6}$ 	&  161.2  $^{+  26.8}_{-  23.4}$ 	&      \tablenotemark{b} \\
                &   -6.8 $\pm$  0.5 &    7.0  $\pm$  0.4 &                       N/A                                	&   87.5  $^{+  27.3}_{-  19.8}$ 	&  105.9  $^{+  30.1}_{-  19.9}$ 	&       \tablenotemark{b} \\
                &   -5.3 $\pm$  0.6 &    3.9  $\pm$  0.4 &   87.7  $^{+  42.9}_{-  29.0}$ 		&   11.4  $^{+   4.9}_{-   6.2}$ 	&   29.9  $^{+   3.4}_{-  25.7}$ 	&      0.011  $\pm$     0.001\\
                &   -3.7 $\pm$  0.7 &    5.6  $\pm$  0.3 &  358.4  $^{+  92.0}_{-  81.6}$ 		&   34.5  $^{+   4.8}_{-   4.5}$ 	&   50.3  $^{+   7.4}_{-   6.3}$ 	&      0.036  $\pm$     0.011 \\
                &   -1.9 $\pm$  0.6 &    4.9  $\pm$  0.2 &  209.0  $^{+  23.6}_{-  28.7}$ 		&   23.9  $^{+   2.6}_{-   2.2}$ 	&   39.1  $^{+   1.9}_{-   1.2}$ 	&      0.20  $\pm$     0.01 \\
                &   -0.5 $\pm$  0.4 &    3.4  $\pm$  0.1 &   50.0  $^{+   9.4}_{-   6.6}$ 		&    2.5  $^{+   3.0}_{-   1.0}$ 	&    4.1  $^{+   0.1}_{-   0.2}$ 	&      1.39  $\pm$     0.02 \\
                &    0.7 $\pm$  0.5 &    3.3  $\pm$  0.1 &   46.4  $^{+   8.7}_{-   3.5}$ 		&     2.7\tablenotemark{c} 		&    2.7\tablenotemark{c} 		&      2.60  $\pm$     0.02 \\
                &    1.7 $\pm$  0.4 &    5.7  $\pm$  0.1 &  392.1  $^{+  64.6}_{-  36.1}$ 		&   36.9  $^{+   2.5}_{-   2.5}$ 	&   52.3  $^{+   5.7}_{-   2.1}$ 	&      1.83  $\pm$     0.06\\
                &    2.8 $\pm$  0.4 &    6.7  $\pm$  0.2 &                      N/A                               	&   75.8  $^{+   9.2}_{-  10.4}$ 	&   94.3  $^{+   8.8}_{-  11.1}$ 	&      1.27  $\pm$     0.37 \\
                &    4.0 $\pm$  0.4 &    5.3  $\pm$  0.2 &  277.0  $^{+  57.6}_{-  39.5}$ 		&   30.0  $^{+   3.4}_{-   3.2}$ 	&   44.0  $^{+   4.7}_{-   2.8}$ 	&      0.025  $\pm$   0.023\tablenotemark{a}  \\
                &    5.4 $\pm$  0.7 &    5.9  $\pm$  0.6 &                       N/A                                	&   40.7  $^{+  20.3}_{-   9.8}$ 	&   60.0  $^{+  19.0}_{-  14.8}$ 	&       \tablenotemark{b} \\
                & & &  & & & \\ 

     09111-1007 &   -3.2 $\pm$  0.6 &    8.8  $\pm$  0.9 &                N/A                         &  243.6  $^{+  95.8}_{-  95.8}$ &  257.6  $^{+  90.4}_{-  90.4}$ &  0.19  $\pm$     0.04 \\
                &   -1.9 $\pm$  0.4 &    7.4  $\pm$  0.4 &                       N/A                                &  112.4  $^{+  21.5}_{-  24.4}$ &  133.5  $^{+  20.9}_{-  27.1}$ &  0.90  $\pm$     0.07 \\
                &   -0.4 $\pm$  0.4 &    6.5  $\pm$  0.2 &                       N/A                                &   64.1  $^{+  11.2}_{-   8.2}$ &   81.6  $^{+  12.3}_{-   7.9}$ &  1.35  $\pm$     0.03 \\
                &    1.1 $\pm$  0.4 &    4.5  $\pm$  0.2 &  150.8  $^{+  21.8}_{-  18.0}$ &   18.6  $^{+   2.2}_{-   2.1}$ &   35.1  $^{+   2.0}_{-   1.6}$ &  0.171  $\pm$     0.005 \\
                &    2.1 $\pm$  0.3 &    6.3  $\pm$  0.2 &                       N/A                                &   55.2  $^{+   8.0}_{-   7.5}$ &   73.0  $^{+   7.6}_{-   7.4}$ &   0.049  $\pm$     0.001 \\
                &    3.1 $\pm$  0.4 &    7.6  $\pm$  0.3 &                       N/A                           &  128.5  $^{+  17.1}_{-  16.3}$ &  149.5  $^{+  15.7}_{-  16.1}$ &  0.001  $\pm$  0.0001 \\
                &    4.2 $\pm$  0.4 &    7.5  $\pm$  0.5 &                       N/A                           &  121.1  $^{+  31.7}_{-  32.5}$ &  142.5  $^{+  29.3}_{-  35.6}$ &  0.0003   $\pm$  0.0001 \\
                &    5.5 $\pm$  0.5 &   10.0  $\pm$  0.8 &                       N/A                                &                       N/A                                &                       N/A                                &   0.00008  $\pm$   0.00005\tablenotemark{a}  \\
                &    7.0 $\pm$  0.6 &   10.0  $\pm$  1 &                       N/A                                &                       N/A                                &                       N/A                                &   0.00001   $\pm$  0.00001\tablenotemark{a}  \\
                & & &  & & & \\ 

     11506+1331 &   -6.7 $\pm$  1.9 &    7.7  $\pm$  1.2 &                       N/A                                &  133.0  $^{+ 131.0}_{-  72.1}$ &  153.7  $^{+ 121.6}_{-  74.9}$ &  0.0073  $\pm$     0.0025 \\
                &   -1.7 $\pm$  1.4 &    \tablenotemark{e} &                       N/A                                & N/A &  N/A   &  0.60  $\pm$     0.01 \\
                &    1.3 $\pm$  1.3 &    6.2  $\pm$  0.2 &                       N/A                                &   51.4  $^{+   8.7}_{-   9.2}$ &   69.2  $^{+   8.8}_{-   8.1}$ &  2.23  $\pm$     0.06 \\
                &    4.4 $\pm$  1.0 &    7.1  $\pm$  0.4 &                       N/A                                &   94.6  $^{+  25.9}_{-  21.2}$ &  111.1  $^{+  30.9}_{-  18.9}$ &  0.053  $\pm$     0.007 \\
                &    7.1 $\pm$  1.1 &    7.7  $\pm$  0.5 &                       N/A                                &  131.1  $^{+  35.5}_{-  31.9}$ &  151.9  $^{+  32.3}_{-  37.2}$ &  0.015  $\pm$     0.001 \\
                &   10.4 $\pm$  1.3 &    6.7  $\pm$  0.4 &                       N/A                                &   72.3  $^{+  21.1}_{-  17.9}$ &   91.3  $^{+  19.0}_{-  19.1}$ &  0.028  $\pm$     0.001 \\
                & & &  & & & \\ 

     10378+1109   &   -4.4 $\pm$  1.7 &    3.6  $\pm$  0.4 &   62.6  $^{+  38.7}_{-  27.8}$ &    6.3  $^{+   6.5}_{-  12.4}$ &    4.3  $^{+  26.3}_{-   0.4}$ &      2.5  $\pm$     1.2 \\
                &   -0.2 $\pm$  1.1 &    \tablenotemark{e} &                       N/A                                & N/A & N/A &      6.73  $\pm$     0.06 \\
                &    2.4 $\pm$  0.9 &    3.9  $\pm$  0.2 &   91.9  $^{+  16.0}_{-  14.6}$ &   11.9  $^{+   2.0}_{-   2.2}$ &   30.2  $^{+   1.0}_{-   1.4}$ &      1.04  $\pm$     0.05 \\
                &    4.7 $\pm$  1.1 &    6.9  $\pm$  0.4 &                       N/A                      &   85.0  $^{+  27.2}_{-  20.8}$ &  103.2  $^{+  30.2}_{-  21.4}$ &      0.37  $\pm$     0.20 \\
                & & &  & & & \\ 

     10565+2448 &   -7.5 $\pm$  0.6 &    7.2  $\pm$  0.2 &                      N/A                              &  100.1  $^{+  14.7}_{-  15.3}$ &  116.1  $^{+  19.9}_{-  13.1}$ &   \tablenotemark{b}  \\
                &   -6.3 $\pm$  0.5 &    8.1  $\pm$  0.2 &                       N/A                              &  158.3  $^{+  11.4}_{-  11.8}$ &  176.7  $^{+  10.3}_{-  10.6}$ &   \tablenotemark{b}  \\
                &   -4.9 $\pm$  0.5 &    6.1  $\pm$  0.6 &  884.4  $^{+   529.4}_{- 529.4}$ &   48.5  $^{+  24.3}_{-  14.2}$ &   66.5  $^{+  25.2}_{-  16.3}$ &    \tablenotemark{b}  \\
                &   -3.4 $\pm$  0.5 &    8.3  $\pm$  0.5 &                       N/A                              &  176.0  $^{+  72.4}_{-  38.6}$ &  192.7  $^{+  69.5}_{-  35.0}$ &   0.010  $\pm$   0.008\tablenotemark{a}  \\
                &   -2.3 $\pm$  0.5 &    6.6  $\pm$  0.3 &                       N/A                                &   65.7  $^{+  16.8}_{-  13.1}$ &   83.6  $^{+  17.0}_{-  13.2}$ &  0.34  $\pm$     0.13 \\
                &   -1.2 $\pm$  0.4 &    4.3  $\pm$  0.2 &  132.2  $^{+  24.0}_{-  22.6}$ &   16.5  $^{+   2.6}_{-   2.3}$ &   33.5  $^{+   2.1}_{-   2.1}$ &  0.42  $\pm$     0.01 \\
                &   -0.1 $\pm$  0.5 &    1.8  $\pm$  0.1 &                       N/A                                & N/A & 2.7\tablenotemark{c} &  7.28  $\pm$     0.08 \\
                &    1.2 $\pm$  0.6 &    5.8  $\pm$  0.2 &  425.7  $^{+ 123.9}_{-  62.6}$ &   38.5  $^{+   5.4}_{-   3.7}$ &   55.9  $^{+   6.6}_{-   5.3}$ &  2.8  $\pm$     0.1 \\
                &    2.2 $\pm$  0.4 &    7.8  $\pm$  0.4 &                       N/A                                &  140.9  $^{+  28.8}_{-  23.8}$ &  160.9  $^{+  26.0}_{-  22.6}$ &   0.14  $\pm$     0.08 \\
\enddata

\tablenotetext{a} {In these apertures, \Hb\ emission was not detected, indicating heavy extinction.  These values represent a lower limit to the SFR based on assuming that the \Hb\ emission line has approximately the same peak as the noise, making it unmeasurable.}

\tablenotetext{b} {\Ha\ emission was not detected in these apertures.}

\tablenotetext{c} {Upper limit of age is presented for very low \EWHB.}

\tablenotetext{d} {These apertures have forbidden/Balmer line ratios that indicate a considerable contribution by AGN.  Star formation rate is excluded due to possible confusion. }

\tablenotetext{e} {The fit provided an unphysically narrow fit of absorption due to contamination from emission, so was rejected from analysis.}

\tablecomments{Col.(1): Projected distance from nucleus to center of measured region in kiloparsecs.  The error corresponds to the spatial width of the aperture.  Col.(2): Absorption equivalent width for the aperture measured in \A\ . Col.(3): Stellar population age in continuing star formation history measured in Myr.  In the apertures where the \EWHB\ is too large for the given stellar population, the section is marked with "N/A".  Col.(4): Stellar population age in Truncated star formation history measured in Myr. Col.(5): Stellar population age in $\delta$ function star formation history measured in Myr. Col.(6): Star formation rate per aperture measured in ${\rm M_{\odot}~yr^{-1}~kpc^{-2}}$.}
\end{deluxetable}

\end{document}